\documentclass[11pt,letterpaper]{article}

\usepackage[T1]{fontenc}
\usepackage[utf8]{inputenc}
\usepackage{lmodern}

\usepackage[margin=1in]{geometry}
\usepackage{microtype}
\usepackage{setspace}
\usepackage{parskip} 
\usepackage{float}

\usepackage[dvipsnames]{xcolor}
\usepackage[colorlinks=true,linkcolor=MidnightBlue,citecolor=ForestGreen,urlcolor=RoyalBlue]{hyperref}

\usepackage{amsmath,amssymb,amsthm}
\usepackage{mathtools}
\usepackage[nameinlink,capitalize]{cleveref}

\usepackage{graphicx}
\usepackage{booktabs}
\usepackage{siunitx}
\usepackage{caption}
\usepackage{tikz}
\usepackage{pgfplots}
\pgfplotsset{compat=1.18}
\usetikzlibrary{positioning,arrows.meta,shapes.geometric,fit,backgrounds,er}

\usepackage{xcolor}
\usepackage{fancyvrb}

\usepackage[backend=bibtex,style=authoryear,maxcitenames=2,maxbibnames=99,doi=true,isbn=false,url=true]{biblatex}

\definecolor{cbgray}{HTML}{999999}
\definecolor{cbteal}{HTML}{1B9E77}
\definecolor{cborange}{HTML}{D95F02}
\definecolor{cbpurple}{HTML}{7570B3}
\definecolor{cbpink}{HTML}{E7298A}
\definecolor{cbsky}{HTML}{66A61E}
\definecolor{cbyellow}{HTML}{E6AB02}
\definecolor{cbbrown}{HTML}{A6761D}
\definecolor{cbblue}{HTML}{1F78B4}
\renewbibmacro*{cite}{%
\printtext[bibhyperref]{%
  \textcolor{green!70!black}{%
  \printnames{labelname}%
  \setunit{\addcomma\space}%
  \printdateextra%
}%
}%
}
\title{Monopoly Deal: A Benchmark Environment for Bounded One-Sided Response Games}
\author{Will Wolf \\
\small \texttt{williamabrwolf@gmail.com}}
\date{\today}

\usepackage{abstract}

\begin{document}

\maketitle

\begin{abstract}
  Card games are widely used to study sequential decision-making under uncertainty, with real-world analogues in negotiation, finance, and cybersecurity.
  Typically, these games fall into three categories based on the flow of control: \emph{strictly-sequential} (where players alternate single actions), \emph{deterministic-response} (where some actions trigger a fixed outcome), and \emph{unbounded reciprocal-response} (where alternating counterplays are permitted).
  A less-explored but strategically rich structure exists: the bounded one-sided response. This dynamic occurs when a player's action briefly transfers control to the opponent, who must satisfy a fixed condition through one or more sequential moves before the turn resolves. We term games featuring this mechanism Bounded One-Sided Response Games (BORGs).

  We introduce a modified version of \emph{Monopoly Deal} as a benchmark environment that specifically isolates the BORG dynamic, where a \emph{Rent} action forces the opponent to sequentially choose payment assets. We demonstrate that the gold-standard algorithm, Counterfactual Regret Minimization (CFR), successfully converges on effective strategies for this domain without requiring novel algorithmic extensions.

  To support efficient, reproducible experimentation, we present a lightweight, full-stack research platform that unifies the environment, a parallelized CFR runtime, and a human-playable web interface, all runnable on a single workstation. This system provides a practical foundation for exploring state representation and policy learning in bounded one-sided response settings.

  The trained CFR agent and source code are available at \url{https://monopolydeal.ai}.
\end{abstract}

\section{Introduction}

In many real-world scenarios, agents must make decisions under uncertainty~\parencite{puterman2014markov}, forming beliefs about hidden information while anticipating the strategies of others~\parencite{osborne1994course}.
An auction is a familiar example: bidders must decide how much to offer without knowing their opponents' private valuations. To study such decision processes in controlled, interpretable domains, researchers often turn to imperfect-information card games with similar strategic elements, including \emph{Kuhn Poker}~\parencite{kuhn1950simplified}, \emph{UNO!}, \emph{Magic: The Gathering}, and \emph{Legends of Runeterra}.
Together, these environments have supported decades of progress in computational game theory and reinforcement learning, from early breakthroughs like solving heads-up limit poker~\parencite{bowling2015heads} to recent superhuman achievements in complex multiplayer domains~\parencite{brown2019superhuman}.

The structure of agent–opponent interaction in these games typically falls into three categories based on the flow of control:
\begin{enumerate}
  \item In \emph{strictly-sequential} games, players alternate single actions. For instance, in \emph{Connect Four}, Alice plays red, then Bob plays black.
  \item In \emph{deterministic-response} games, some actions trigger a fixed, rule-based outcome. For example, if Alice plays \emph{Draw Two} in \emph{UNO!}, Bob draws two cards.
  \item In \emph{unbounded reciprocal-response} games, certain actions allow alternating counterplays before the turn resolves. For instance, in \emph{Magic: The Gathering}, Alice casts a spell, Bob plays a counterspell, and Alice may respond again, continuing until both pass.
\end{enumerate}

While these interaction patterns capture most well-studied card games, they do not fully represent a separate, common structure: where an action briefly transfers control to an opponent who must take a short, non-reciprocal sequence of actions to satisfy a fixed condition before play resumes.
Such response phases are ubiquitous in structured real-world interactions, such as time-sensitive options trading or regulatory compliance workflows, where one party initiates a request, and the opponent must respond with a bounded, non-reciprocal sequence of actions (e.g., settling a margin call or invoking a cure period) before the main transaction resumes. We refer to games exhibiting this structure as \textbf{Bounded One-Sided Response Games (BORGs)}.

To study this interaction pattern in a compact and reproducible form, we present a modified version of \emph{Monopoly Deal} that isolates the BORG dynamic while remaining compatible with standard extensive-form representations.
Each player takes turns playing one or more cards, choosing among actions such as acquiring properties, collecting cash, or charging rent based on owned property sets.
When a \emph{Rent} card is played, control temporarily transfers to the opponent, who must respond by selecting a sequence of \emph{Cash} or \emph{Property} cards to satisfy the owed amount, or by canceling the demand with a \emph{Just Say No} card.
Once the debt is resolved, the turn concludes—forming a bounded, one-sided response phase that is distinct from standard deterministic or reciprocal-response models.

We apply Counterfactual Regret Minimization (CFR)~\parencite{zinkevich2007regret,lanctot2009monte,brown2019solving} to this environment using a compact state representation that accommodates the response phases. This demonstrates that established regret-minimization techniques can be applied directly to BORGs and converge reliably, without requiring novel algorithmic extensions.

To support reproducible experimentation, we develop a lightweight, full-stack research platform that unifies the game environment, a locally parallelized CFR runtime, and a human-playable web interface—all runnable on a single workstation.
The system emphasizes accessibility and introspection: users can launch and monitor training runs, inspect intermediate states and policies, and easily interact with learned models.
This framework lowers the barrier to entry for studying BORGs.

\paragraph{Contributions.}
This work makes three primary contributions:

\begin{enumerate}
  \item Formalizes the BORG dynamic within a modified \emph{Monopoly Deal} environment, providing a reproducible benchmark for studying bounded one-sided response games.
  \item Demonstrates the tractability of BORGs by showing that existing Counterfactual Regret Minimization (CFR) techniques apply directly and converge efficiently in this setting.
  \item Presents a lightweight research platform that unifies the \emph{Monopoly Deal} environment, a parallelized CFR runtime, and a web interface for accessible, reproducible experimentation.
\end{enumerate}

\section{Background and Notation}
\label{sec:background}

We model our environment as a two-player, zero-sum, imperfect-information game represented in the \emph{Extensive-Form Game (EFG)} setting~\parencite{osborne1994course,kuhn1953extensive} equivalent to a partially observable stochastic game~\parencite{kaelbling1998planning}. The game is defined by the tuple
\[
  \mathcal{G} = (\mathcal{S}, \mathcal{Z}, \mathcal{A}, P, u, \mathcal{I}),
\]
where $\mathcal{S}$ denotes the set of all reachable \emph{game states} (or histories/nodes in the game tree), $\mathcal{Z} \subseteq \mathcal{S}$ the set of terminal states, and $\mathcal{A}(s)$ the set of actions available at state $s$. $P(s)$ is the player function indicating whose turn it is, and $u_i: \mathcal{Z} \to \mathbb{R}$ is the utility (reward) for player $i$. The game is strictly zero-sum: $u_1(z) = -u_2(z)$.

Each player $i \in \{1,2\}$ makes decisions based on their \emph{information set $\mathcal{I}_i$}, which is a partition of the nodes in $\mathcal{S}$ such that all states $s \in \mathcal{I}_i$ are indistinguishable to player $i$ based on their private observations. The \emph{policy $\boldsymbol{\sigma}_i$} is a probability distribution over actions conditioned on the current information set $\mathcal{I}_i$.

\subsection*{Bounded One-Sided Response Operator}

We restrict our analysis and subsequent environment implementation to the two-player instance of \emph{Monopoly Deal}.

To explicitly model the BORG dynamic within the EFG framework, we introduce a \textbf{bounded-response operator $\boldsymbol{\varrho}$} that captures temporary, one-sided control transfers between players. For any state–action pair $(s,a_i)$ that triggers a response,
\[
  \boldsymbol{\varrho}(s,a_i) \subseteq \mathcal{S}
\]
denotes the finite subgraph of states reachable during the opponent's bounded response phase. Within this subgraph, only the opponent $j \neq i$ may act, producing a sequence
$\{a_j^1, a_j^2, \ldots, a_j^k\}$ until a stopping condition $\tau(s') = \text{True}$ is met.
The system then transitions deterministically to the post-response state $s''$ (where the BORG phase results are applied and the turn is concluded) and returns control to the next player in turn order.

This structure defines a piecewise-alternating control pattern that generalizes strictly sequential games. Unlike deterministic-response games, which trigger a fixed, rule-based outcome, or unbounded reciprocal-response games, which allow alternating counterplays of arbitrary depth, BORGs introduce a non-reciprocal subphase in which one player acts repeatedly until a fixed condition is satisfied. This formulation highlights a common but underexplored pattern: finite, one-sided sequences of actions governed by a fixed termination rule.

\subsection*{Counterfactual Regret Minimization}

Counterfactual Regret Minimization (CFR)~\parencite{zinkevich2007regret} is a regret-based method for computing approximate Nash equilibria in sequential, imperfect-information games.

Let $\sigma = (\sigma_1, \sigma_2)$ denote the joint strategy profile,
and let $\pi_i^\sigma(s)$ denote the contribution of player $i$'s actions to the probability of reaching state $s$ under the joint policy $\sigma$.
We define the \emph{counterfactual reach probability} of state $s$ for player $i$ as
\[
  \pi_{-i}^\sigma(s) = \prod_{j \neq i} \prod_{a_t \in \text{path}(s)} \sigma_j(a_t | I_t),
\]
the product of the action probabilities of all other players (and chance) along the path from the root to $s$.

Each player maintains an information set policy $\sigma_i(I)$ over actions $a \in \mathcal{A}_i(I)$, where $I$ denotes the set of states sharing the same observation sequence.
The \emph{counterfactual value} of taking action $a$ at information set $I$ under policy $\sigma$ is
\[
  v_i^\sigma(I,a)
  = \sum_{z \in Z_I} \pi_{-i}^\sigma(z[I,a]) \, u_i(z),
\]
where $Z_I$ denotes the set of terminal states $z$ reachable after taking action $a$ at information set $I$.
Equivalently, $v_i^\sigma(I,a)$ is the average utility to player $i$ over all terminal states $z$ reachable after taking $a$ at $I$ weighted by the reach probability of all other players (and chance).

The \emph{expected utility} for player~$i$ at information set~$I$ under policy $\sigma_i$ is then
\[
  v_i^\sigma(I) = \mathbb{E}_{a \sim \sigma_i(\cdot|I)}[v_i^\sigma(I,a)]
  = \sum_{a \in \mathcal{A}_i(I)} \sigma_i(a|I) \, v_i^\sigma(I,a).
\]
At each iteration~$t$, the \emph{instantaneous regret} for each action $a \in \mathcal{A}_i(I)$ is accumulated as:
\[
  r_i^t(I,a) = v_i^{\sigma^t}(I,a) - v_i^{\sigma^t}(I),
  \qquad
  R_i^T(I,a) = \sum_{t=1}^{T} r_i^t(I,a),
\]
and policies are updated via regret matching:
\[
  \sigma_i^{T+1}(I,a) =
  \begin{cases}
    \dfrac{\max(R_i^T(I,a), 0)}{\sum_{a'} \max(R_i^T(I,a'), 0)} &
    \text{if denominator} > 0,                                                      \\[1.2ex]
    \dfrac{1}{|\mathcal{A}_i(I)|}                               & \text{otherwise.}
  \end{cases}
\]

Convergence is typically measured by the \emph{maximum expected regret (MER)}. For an imperfect-information game, CFR minimizes this metric to guarantee that the average strategy approaches a \emph{Nash equilibrium} as the number of iterations increases~\parencite{zinkevich2007regret,lanctot2009monte}. The MER is computed as the counterfactual-reach-weighted average of the maximum instantaneous regret across all visited information sets:
\[
  \text{MER} = \frac{\sum_{I} \max_{a \in \mathcal{A}_i(I)} R_i(I, a) \cdot \pi_{-i}(I)}{\sum_{I} \pi_{-i}(I)}
\]
As the average regret $\bar R_i^T$ approaches zero,
the average strategy $\bar\sigma_i^T$ converges to a Nash equilibrium.

\textbf{Monte Carlo CFR.}
In large-scale games, the full tree traversal required by vanilla CFR is computationally infeasible. To address this, \textbf{Monte Carlo Counterfactual Regret Minimization (MCCFR)} \parencite{lanctot2009monte} was introduced as a general family of domain-independent algorithms that uses sampling to estimate the expected utilities and accumulated regrets. The core theoretical result is that MCCFR performs the same regret updates as the full-traversal CFR in expectation, thus maintaining the convergence guarantees. It encompasses various sampling schemes, including outcome sampling and external sampling.

\paragraph{External Sampling.}
In our implementation, we utilize the \textbf{External Sampling (ES-CFR)} variant \parencite{lanctot2009monte} of MCCFR, which samples only external factors (opponent and chance actions) during its recursive traversal. However, different from the standard approach, our implementation uses \emph{action-based rollouts}: at the specific information set $I$ where regret is being updated, all actions $a \in \mathcal{A}_i(I)$ are enumerated, and all subsequent actions by all players are single-sampled for $N$ full-game trajectories. Furthermore, to reduce variance during training, our implementation omits the counterfactual reach probability $\pi_{-i}^\sigma(s)$ from the instantaneous regret calculation, relying instead on the unweighted average utility derived from the rollouts.

\section{Modified Monopoly Deal Environment}

We introduce a two-player, zero-sum version of \emph{Monopoly Deal}~\parencite{hasbromonopolydealgame,hasbromonopolydealrules} that preserves the strategic structure of the original game while simplifying its rules to isolate bounded one-sided response dynamics. Each player seeks to complete colored property sets by acquiring properties, collecting rent, and managing cash resources under imperfect-information.

\paragraph{Card Types.}
The deck contains four categories of cards.
\emph{Property} cards come in three colors—Brown, Green, and Pink—with fixed rent progressions and associated cash values.
\emph{Cash} cards serve purely as currency for settling rent, and thus a ``buffer'' against paying rent with acquired properties.
\emph{Rent} cards (colored Brown, Green, and Pink) allow a player to charge rent proportional to the number of owned properties (\#) of the \emph{Rent} card's color, triggering a response phase if $\# > 0$. For instance, rent for Green properties is charged in amounts $\$2, \$4$, and $\$7$ when the renting player owns 1, 2, and 3 Green properties, respectively.
  \emph{Just Say No} cards cancel a rent demand entirely.

  \paragraph{Turns and Streaks.}
  A turn is defined by the initiating player's main action (e.g., playing a card or passing). Crucially, a single turn encompasses the entire resolution of that action, including any subsequent bounded response phase ($\varrho$). Thus, one game turn may contain multiple node transitions within the EFG tree. Consecutive turns taken by the same player form a \emph{streak}, which continues until the player has completed a fixed number (e.g. 2) of turns (or passes voluntarily). When a streak ends, control transfers to the opponent, initiating her next turn. Players alternate streaks throughout the game.

  \paragraph{Hand Size.}
  Players are initially dealt five cards, and select two new cards from the deck at the start of their streak. In the original game, players are required to discard cards above seven at the end of their turn. For simplicity, we remove this rule in our implementation.

  \paragraph{Bounded One-Sided Response.}
  When a \emph{Rent} card is played, control transfers temporarily to the opponent under the response operator~$\varrho$.
  The opponent must satisfy a fixed rent amount by paying with cash or property, canceling with a \emph{Just Say No} card, or ``yielding'' if unable to pay.

  In the original game, the initiating player may cancel a \emph{Just Say No} card with another \emph{Just Say No} card. To maintain the BORG structure, we disallow this in our implementation.

  Only the responding player acts during this phase, and play deterministically resumes once the rent is settled or blocked.
  This produces a bounded, non-reentrant subphase distinct from reciprocal-response games, where control may alternate repeatedly.

  \paragraph{Objective and Termination.}
  A player wins upon completing a fixed number of full property sets.
  If the deck is exhausted and neither player meets this condition, the game ends in a draw.
  Compared to the commercial game, this version reduces property types, removes most action cards (e.g. \emph{Sly Deal}, \emph{Forced Deal}), and limits cash denominations.
  These simplifications preserve strategic depth while yielding a tractable, reproducible testbed for studying BORGs.

  \section{Related Work}

  Our work sits at the intersection of three areas of sequential decision-making research: 1) Benchmark environments for studying specific game dynamics, 2) The demonstration of algorithmic tractability, and 3) The creation of lightweight, reproducible research platforms.

  \paragraph{Prior Research on Monopoly Deal.}
  Previous research has explored implementing game-playing algorithms for the original version of \emph{Monopoly Deal}. This work used heuristic-driven player personalities (e.g., Aggressive, Defensive), hand-coded strategic priorities and graph-traversal techniques such as Breadth-First Search (BFS) to achieve varied and competitive play~\parencite{Lazarusli2015Monopoly}. Our work differs by employing Counterfactual Regret Minimization (CFR) to derive a robust policy through intent-based state abstraction and self-play alone. To measure the competitiveness of the CFR policy, we use similar heuristic-based agents as baseline opponents.

  \paragraph{Existing Benchmark Environments.}
  The community currently benefits from numerous platforms for studying diverse games. \textbf{OpenSpiel}~\parencite{lanctot2019openspiel} provides a unified interface for a wide collection of board, card, and stochastic games together with standard algorithms for self-play learning. \textbf{RLCard}~\parencite{zha2020rlcard} offers a lightweight toolkit focused on card games such as \emph{Texas Hold'em} and \emph{UNO!}, with a standardized API and strong reproducibility guarantees. \textbf{PettingZoo}~\parencite{terry2021pettingzoo} generalizes the OpenAI Gym interface~\parencite{brockman2016gym} to multi-agent settings, introducing the Agent Environment Cycle abstraction for turn-based play. Similarly, \textbf{Hanabi Learning Environment}~\parencite{bard2020hanabi} offers a compact, domain-specific benchmark for cooperative reasoning. Together, these frameworks emphasize modularity and standardized interfaces for broad experimentation in imperfect-information games. Our environment adds a new, specific domain—BORGs, represented by a modified version of \emph{Monopoly Deal}—to this growing collection.

  \paragraph{Scaling Algorithms for Complex Games.}
  To achieve superhuman performance in the most complex imperfect-information domains, such as \emph{No Limit Texas Hold 'Em}, researchers have developed powerful learning algorithms and large-scale training frameworks. Early breakthroughs like \textbf{DeepStack}~\parencite{moravcik2017deepstack} and subsequent systems like \textbf{Libratus} and \textbf{Pluribus}~\parencite{brown2019superhuman} required scaling Counterfactual Regret Minimization (CFR) to handle state spaces too large for tabular methods. To overcome this challenge, they introduced \emph{deep function approximation} techniques for generalizing across states, leading to modern neural models like \textbf{Deep CFR}~\parencite{brown2018deepcfr} and \textbf{ReBeL}~\parencite{brown2020rebel}, as well as the tooling and infrastructure (e.g. distributed GPU clusters) necessary to train these systems. This class of approaches represents the state-of-the-art for tackling domains where tabular CFR is intractable. In contrast, in order to begin studying the BORG dynamic (in an abstracted state space of modest size), our work uses classic tabular CFR alone to achieve competitive performance against baseline models.

  \paragraph{Lightweight and Reproducible Research Platforms.}
  \begin{sloppypar}
    Different from research that relies on large-scale distributed systems, a parallel tradition emphasizes high-throughput, reproducible training with efficient local execution. This philosophy underlies systems such as \textbf{ELF}~\parencite{tian2017elf}, which introduced a flexible C++/Python framework designed for high-efficiency simulation and fast local training, often achieving breakthrough performance on single-workstation hardware. Other efforts such as \textbf{Sample Factory}~\parencite{petrenko2020samplefactory} and \textbf{SLM Lab}~\parencite{kenggraesser2017slmlab} adopt modular, configuration-driven designs that facilitate local parallelism and consistent experiment logging on a single workstation. Furthermore, reproducibility itself has become a central theme in reinforcement-learning research, with works such as~\parencite{henderson2017deeprlmatters} highlighting the importance of seeding, implementation variance, and robust evaluation protocols. Our work follows directly in this tradition: we provide a lightweight, full-stack research platform designed for single-workstation execution, allowing for fast and transparent experimentation in the bounded-response domain.
  \end{sloppypar}

  \section{System Design and Architecture}
  \label{sec:system}

  \subsection{Design Goals}

  The system is a lightweight, full-stack research platform for studying bounded one-sided response games through \emph{Monopoly Deal}.
  Its design emphasizes reproducibility, clarity, and accessibility over raw throughput.
  The following goals guided the architecture.

  \begin{itemize}
    \item \textbf{Fast convergence.}
          The platform achieves stable Counterfactual Regret Minimization convergence within twenty minutes of wall-clock training time on a single workstation.
          This efficiency results from the interaction between a compact state abstraction and an efficient parallel CFR runtime.

    \item \textbf{Introspection and Logging.}
          All training metrics and checkpoints are logged continuously to Weights \& Biases. This enables researchers to monitor convergence, inspect policy updates, and audit experiment provenance.

    \item \textbf{Human interaction.}
          Trained agents can be easily loaded into a web interface for interactive play, shortening the loop between quantitative training results and qualitative behavioral evaluation.

    \item \textbf{Reproducibility.}
          Deterministic random seeds, step-indexed checkpoints, and embedded commit hashes allow training runs to be resumed or replayed exactly. This emphasis aligns with broader community efforts toward reliable reinforcement-learning evaluation~\parencite{henderson2017deeprlmatters}.

    \item \textbf{State Representation.}
          The internal data model cleanly encodes hidden information, bounded response contexts, and action legality.
          This representation supports not only CFR but future algorithms requiring precise serialization of the game's information sets.
  \end{itemize}

  \noindent
  Together, these goals motivated a modular, end-to-end design combining a transparent environment, a parallelized runtime, and an accessible interface—making high-fidelity experimentation feasible on a single workstation.

  \subsection{Game Environment}
  The environment uses a simplified two-player variant of \emph{Monopoly Deal} with a reduced card set, which allows us to focus on the BORG dynamic without the complexity of the original game. Furthermore, the number of property types and required sets are restricted to keep the number of unique information sets manageable. This allows state abstraction techniques to begin from a simpler starting point, and also allows the memory footprint of the regret and policy lookup tables to be kept within single-workstation RAM.

  \subsection{Architecture Overview}

  The system comprises two principal stacks—a training stack for self-play learning and a serving stack for human–model interaction—connected through a shared JSON checkpoint artifact.

  The training stack is launched through a single command that configures the CFR learner and dispatches self-play games across local Ray workers. A central process maintains the global policy, regret, and reach-probability managers, while each worker executes a complete self-play game using the current snapshot. When a game finishes, the learner aggregates regret and policy updates synchronously into a unified global state. Metrics are logged to Weights \& Biases, and final checkpoints are serialized as human-readable JSON snapshots containing configuration and learner states.

  The serving stack loads trained checkpoints into a FastAPI backend, exposing endpoints that utilize the same game engine used in training. A React/Next.js frontend provides an interactive web interface for human–model play and state visualization. All game metadata and move histories are persisted to a PostgreSQL database for behavioral auditing and fault tolerance (allowing the game to be resumed exactly should the server fail). The stack supports three execution modes—local development, local containerized development, and production deployment on Google Cloud Run to \href{https://monopolydeal.ai}{https://monopolydeal.ai}.

  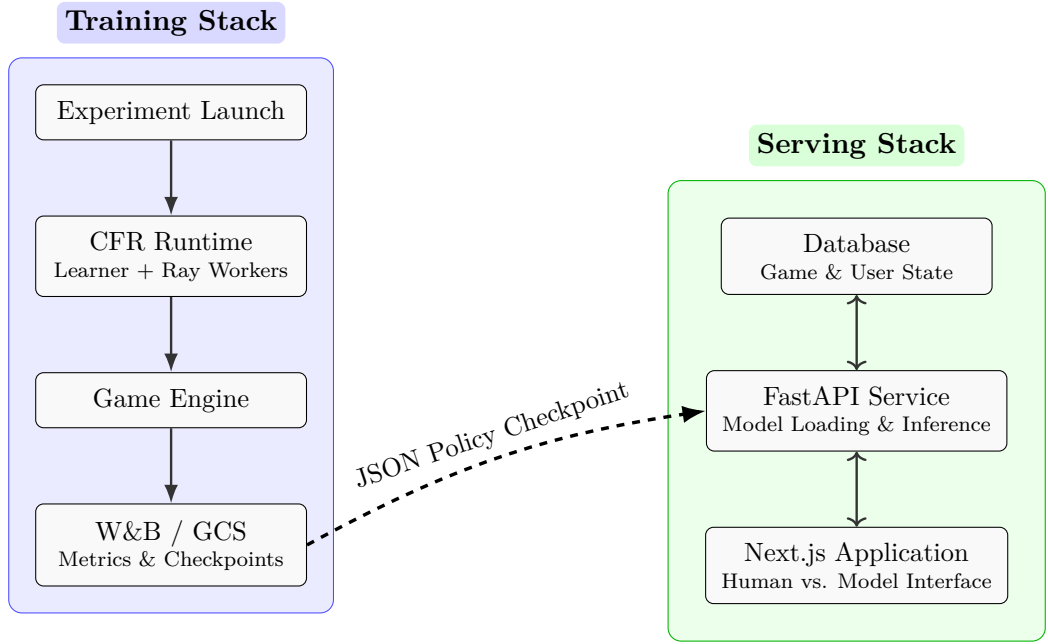
\begin{figure}[t]
    \centering
    \begin{tikzpicture}[
        font=\small,
        node distance=1.0cm,
        box/.style={
            draw, rounded corners=3pt, fill=gray!5,
            minimum width=3.6cm,
            align=center,
            inner sep=6pt
          },
        flow_arrow/.style={-Latex, thick, rounded corners, line width=0.9pt, color=gray!40!black},
        bidirectional_arrow/.style={<->, thick, rounded corners, line width=0.9pt, color=gray!40!black}, 
        data_arrow/.style={-Latex, dashed, rounded corners, line width=1.3pt, color=black},
      ]

      \node[box] (cli) {Experiment Launch};
      \node[box, below=of cli] (runtime) {CFR Runtime\\[-2pt]\scriptsize Learner + Ray Workers};

      \node[box, below=of runtime] (engine) {Game Engine};

      \node[box, below=of engine] (wandb) {W\&B / GCS\\[-2pt]\scriptsize Metrics \& Checkpoints};

      \foreach \a/\b in {cli/runtime, runtime/engine, engine/wandb}
      \draw[flow_arrow] (\a) -- (\b);

      \begin{scope}[on background layer]
        \node[
          draw=blue!70, rounded corners=5pt, inner sep=10pt, fill=blue!8!white,
          fit=(cli)(wandb)
        ] (trainbox) {};
      \end{scope}
      \node[font=\bfseries, fill=blue!15, rounded corners=3pt, inner sep=3pt, above=5pt of trainbox.north] {Training Stack};

      \node[box, right=5.5cm of runtime] (db) {Database\\[-2pt]\scriptsize Game \& User State};
      \node[box, below=of db] (fastapi) {FastAPI Service\\[-2pt]\scriptsize Model Loading \& Inference};
      \node[box, below=of fastapi] (nextjs) {Next.js Application\\[-2pt]\scriptsize Human vs. Model Interface};

      \draw[bidirectional_arrow] (db) -- (fastapi);
      \draw[bidirectional_arrow] (fastapi) -- (nextjs);

      \begin{scope}[on background layer]
        \node[
          draw=green!70!black, rounded corners=5pt, inner sep=14pt, fill=green!8!white,
          fit=(db)(nextjs)
        ] (servebox) {};
      \end{scope}
      \node[font=\bfseries, fill=green!15, rounded corners=3pt, inner sep=3pt, above=5pt of servebox.north] {Serving Stack};

      \draw[data_arrow, bend left=10]
      (wandb.east) to node[midway, above, sloped, font=\small, color=black] {JSON Policy Checkpoint} (fastapi.west);

    \end{tikzpicture}

    \caption{System architecture: training and serving stacks. The training stack runs CFR self-play experiments and logs metrics and checkpoints. The trained policy is exported as JSON to the serving stack, where it is loaded into a FastAPI service backed by a database and accessed through a Next.js frontend.}
  \end{figure}

  \subsection{Data Model and State Representation}

  Game state is modeled through a hierarchy of immutable, serializable data-classes that cleanly separate public and private information under imperfect observability. The top-level \texttt{GameState} represents a player's information set—the view available to that player—rather than a full world state. It comprises three types of information:

  \begin{sloppypar}
    \begin{itemize}\raggedright
      \item \textbf{Player and opponent state} is captured by \texttt{PlayerState} and \texttt{OpponentState} objects. \texttt{PlayerState} stores the acting player's private hand and public holdings, while \texttt{OpponentState} contains only public holdings. These holdings are partitioned into distinct ``piles'': \texttt{Hand}, \texttt{CashPile}, and \texttt{PropertyPile}. This asymmetry ensures that each \texttt{GameState} instance encodes a legal information set from the acting player's perspective.

      \item \textbf{Bounded One-Sided Response Game (BORG) Context} is contained within the \texttt{TurnState}, which tracks the \texttt{turn\_idx}—the global counter of turns played—and \texttt{streak\_idx}—the counter of turns taken by the initiating player in a given streak. The BORG dynamic is defined by an optional \texttt{response\_ctx}, which, when present, represents the deterministic, finite subgraph of the response phase. This context records three elements:
            \begin{enumerate}
              \item The \textbf{initiating action} (e.g., \emph{Rent}) that triggered the response.
              \item The \textbf{pre-response state} of the initiating player.
              \item The sequence of \textbf{response actions} taken by the opponent so far.
            \end{enumerate}
            The BORG structure is formally enforced by the action's \textbf{response definition}, e.g. \texttt{RentCardResponseDefinition}. This interface is designed for extensibility; to implement a new BORG action, the definition must provide methods to govern the phase's logic:
            \begin{enumerate}
              \item Check if the phase is necessary (\texttt{response\_required}).
              \item Define the valid action set at any point (\texttt{get\_valid\_responses}).
              \item Define the precise termination condition of the phase (\texttt{response\_complete}).
            \end{enumerate}

      \item \textbf{Configuration and metadata} includes the \texttt{GameConfig} (which fixes ruleset parameters) and the \texttt{random\_seed} (which supports deterministic play).
    \end{itemize}
  \end{sloppypar}

  All components implement a shared \texttt{Serializable} interface. Frozen data-classes enforce immutability, and a deterministic hash of the serialized encoding serves as the information-set key, enabling consistent indexing and checkpoint recovery.

  \begin{sloppypar}\raggedright
    Each \texttt{GameState} also references the \texttt{abstraction\_cls} and \texttt{resolver\_cls}. The default \texttt{IntentStateAbstraction} maps concrete \emph{actions} (e.g., playing a Green \emph{Property} card) to intent-based \emph{abstract actions} (e.g., \textsc{StartNewPropertySet}) suitable for policy learning. The \texttt{GreedyActionResolver} deterministically deduplicates equivalent choices (e.g., selecting the highest-value cash card when multiple options exist for a \texttt{Cash} abstract action selected from the player's policy).
  \end{sloppypar}

  \subsection{Runtime and Reproducibility Model}

  Training runs are orchestrated through a central $\text{Counterfactual Regret Minimization (CFR)}$ learner that dispatches self-play games to local Ray workers. The execution model supports three distinct parallelism strategies, managed by a \texttt{ParallelismStrategy} class, which balance runtime efficiency against state consistency and determinism:
  \begin{enumerate}
    \item \textbf{Sequential ($\texttt{none}$)}: The learner runs one self-play game at a time. Updates are applied synchronously after each game, ensuring strict game-order consistency and maximum determinism, but with no parallel speedup.
    \item \textbf{Parallel Unordered Update ($\texttt{parallel-unordered-update}$)}: Workers launch games concurrently and return results asynchronously. The learner applies updates (merging regret and policy deltas) immediately upon worker completion without regard to game index. This mode is the fastest, allowing for approximately 20 minutes of wall-clock training time for convergence, but it is not deterministic.
    \item \textbf{Parallel Batch Ordered Update ($\texttt{parallel-batch-ordered-update}$)}: This is the default mode and is fully deterministic. Workers launch games in batches (sized by the number of CPUs). The learner waits for the entire batch to complete, then aggregates and applies all updates synchronously in ascending game-index order before launching the next batch. This ensures state consistency and full determinism but increases the wall-clock training time required for convergence.
  \end{enumerate}

  \paragraph{Random Seeding}
  Reproducibility is supported via $\text{\emph{random seeds}}$: a single run-level seed initializes pseudo-random number generators, and each individual game's trajectory is seeded deterministically using a function of the primary seed and the game index.

  \paragraph{Global State Management}
  A single learner process maintains the global policy, regret, and reach-probability managers as shared objects. Each worker simulates a complete self-play game using a snapshot of the current global state. For the parallel modes, the learner ensures state consistency by putting the updated CFR object into Ray's object store after each update sequence (game or batch), making the new state available for subsequent jobs.

  \paragraph{Checkpointing}
$\text{State persistence}$ is managed by writing a checkpoint containing the complete learner state (policy buffers, regret tables, etc.) and full configuration metadata (ruleset, abstraction class names, and seeds). Checkpoints are written according to a configurable interval, with the default being only at the final game completion. The system supports $\text{mid-run resumption}$ by synchronously loading the highest indexed checkpoint found in the specified checkpoint store. An example checkpoint can be found in Section~\ref{sec:checkpoint} of the appendix.

  \paragraph{Logging}
  At launch, the system (optionally) logs all experiment configuration arguments to $\text{Weights \& Biases (W\&B)}$ as experiment metadata. The learner then logs expected regret statistics, win rates against baseline opponents, information-set update quantiles, and median policy probabilities. The game index serves as the canonical step for visualization and logging, ensuring metric uniqueness and consistency even if worker completions occur out-of-order (as is possible in the unordered parallel mode).

  \subsection{Human Interaction and Policy Evaluation}

  Trained policies can be inspected and played against directly through an integrated web interface built on a FastAPI backend and a Next.js frontend. The same \texttt{Game} engine used for training is exposed as a FastAPI service that loads a serialized policy checkpoint and executes actions through identical environment logic. The frontend connects via API endpoints and renders each state transition in real time, allowing human players to compete against trained CFR agents.

  The interface supports detailed state inspection at each step, including the agent's policy probabilities and information-set visit counts, and adjustable agent pacing to allow a human player to follow the CFR agent's decision flow. All game interactions are persisted to a PostgreSQL database, including action histories, game outcomes, and player metadata, allowing for post-hoc analysis of completed games. Because the serving stack reuses the same deterministic engine and serialization pathway as training, every human–model game can be reconstructed exactly from its stored log.

  The coupling between training and interaction layers elevates the system from a headless experiment runner to a unified research platform, enabling policies to be trained, evaluated, visualized, and examined within a single environment. A screenshot of the web interface can be found in Section~\ref{sec:ui_screenshot} of the appendix.

  \section{Methods}
  \label{sec:methods}

  \subsection{Monte Carlo CFR with Variance Reduction}
  \label{sec:mccfr}

  Our implementation follows the tabular Monte Carlo Counterfactual Regret Minimization (MCCFR) variant. Instead of utilizing the standard recursive traversal of External Sampling (ES-CFR), we employ a custom, non-recursive, \emph{action-based rollout strategy}. This approach, combined with a compact state abstraction, enables rapid convergence for our small-scale environment.

  At the target information set $I$ for player $i$, all available actions $\mathcal{A}_i(I)$ are enumerated. For each action $a$, we execute $N$ full-game trajectories to estimate the action's counterfactual value. During these rollouts, all subsequent actions—by the target player $i$, opponent $j$, and chance—are single-sampled from their respective policies. The opponent $j$ uses the historical average policy $\bar{\sigma}_j(\cdot|I)$, and the target player $i$ uses the latest, exploration-enabled policy $\sigma_i(\cdot|I)$.

  \paragraph{Intent-based abstraction.}
  The state space is compressed through a minimal \emph{intent-based abstraction} that represents only the set of available abstract actions and the streak index at each decision point.
  Formally, each concrete information set $I$ is mapped to a reduced representation
  \[
    \phi(I) = (\texttt{actions}(I), \texttt{streak\_idx}),
  \]
  where $\texttt{actions}(I)$ is the sorted tuple of abstract actions available to the acting player.
  The action vocabulary includes
  \textsc{StartNewPropertySet}, \textsc{AddToPropertySet}, \textsc{CompletePropertySet},
  \textsc{Cash}, \textsc{AttemptCollectRent}, \textsc{JustSayNo},
  \textsc{GiveOpponentCash}, \textsc{GiveOpponentProperty},
  \textsc{Pass}, \textsc{Yield}, and \textsc{Other}.
  No additional quantitative features—such as hand size, property counts, or opponent statistics—are encoded.
  This compact representation yields roughly one hundred unique information sets in practice, which produces competitive strategies with minimal memory overhead and fast convergence.

  Because the abstraction key is composed only of the current set of available abstract actions and the streak index, a state within the BORG phase is differentiated from a non-response state solely by the unique, restricted set of valid actions (e.g., \textsc{GiveOpponentCash}, \textsc{JustSayNo}) enforced by the response operator $\boldsymbol{\varrho}$. This implicit differentiation allows us to use Counterfactual Regret Minimization on BORGs without modifying the algorithm itself.

  \paragraph{Learning procedure.}
  At each iteration~$t$, the learner samples full trajectories through self-play.
  For each visited information set~$I$ with available actions~$\mathcal{A}_i(I)$, we estimate the counterfactual value of action $a$ via unweighted Monte Carlo averaging:
  \[
    \hat v_i^{\sigma^t}(I,a) = \frac{1}{N}\sum_{n=1}^{N} u_i(z_n|I,a),
  \]
  where $z_n$ denotes a terminal outcome sampled after taking~$a$ at~$I$ and following the current policy thereafter. This average serves as the estimator for the true counterfactual value $v_i^{\sigma^t}(I,a)$. Critically, we omit the counterfactual reach probability weighting $\pi_{-i}^\sigma(z)$ that appears in the standard ES-CFR definition. While this introduces a small bias relative to the exact MCCFR estimator, it reduces variance while still producing competitive, fast-converging policies.

  The instantaneous regret $r_i^t(I,a)$ is computed based on the estimated values:
  \[
    r_i^t(I,a) = \hat v_i^{\sigma^t}(I,a) - v_i^{\sigma^t}(I),
  \]
  where $v_i^{\sigma^t}(I) = \sum_{a' \in \mathcal{A}_i(I)} \sigma_i(a'|I) \, \hat v_i^{\sigma^t}(I,a')$. The regret is accumulated over time as $R_i^T(I,a) \leftarrow R_i^T(I,a) + r_i^t(I,a)$, and the policy $\sigma_i^{T+1}(I,a)$ is updated using regret matching as defined in Section ~\ref{sec:background}.

  \paragraph{Stabilization heuristics.}
  Two practical simplifications further improve numerical stability and model behavior:
  \begin{itemize}
    \item \textbf{Action Clamping for Progress.} Card-playing actions are essential for advancing the game state toward a win, whereas non-card actions (e.g. \textsc{Pass}) represent passive delay. To ensure the agent prioritizes immediate progress, we modify the regret matching calculation: if any card-playing action holds positive accumulated regret, the regret of all non-card actions is clamped to $\leq 0$, ensuring that the agent only chooses passive delay when it is the only action with positive regret. The regret for the \textsc{Other} action, which represents an action that does not correspond to a meaningful high-level intent given the game state, is similarly suppressed.
    \item \textbf{Exploration.} During training, actions are sampled $\varepsilon$-greedily with $\varepsilon = 0.1$, fixed across all iterations, which encourages the policy to visit all available information sets.
  \end{itemize}

  \paragraph{Policy averaging.}
  Following standard CFR practice, the Nash Equilibrium solution is derived from the average policy, $\bar\sigma_i$. For practical deployment, we maintain a \emph{fixed-size buffer} of recent strategies to compute the average policy~$\bar\sigma_i$ used for evaluation.
  The buffer size is configurable (default $N=10$), and the average is taken uniformly over stored policies.
  This produces a stable, low-variance policy estimate for deployment while keeping a constant memory footprint.

  \section{Experiments}
  \label{sec:experiments}

  We evaluate the system's ability to learn stable, high-performing policies in bounded one-sided response environments through self-play on our modified \emph{Monopoly Deal} environment.
  Experiments focus on convergence speed, policy stability, and interpretability of learned strategies under the intent-based abstraction introduced in Section~\ref{sec:methods}.
  All experiments were conducted on a single Apple~M1 workstation with 8~logical cores and 16~GB~RAM, achieving convergence in approximately 19 minutes of wall-clock time using the \texttt{parallel-unordered-update} parallelism mode.

  \subsection{Experimental Setup}

  Training uses Monte Carlo CFR with action-based rollouts and a fixed $\epsilon$-greedy exploration rate.
  As detailed in Section~\ref{sec:methods}, this approach samples opponent/chance actions using the current policy.
  Each self-play game proceeds until termination or a cap of $250$ turns, with policy and regret updates applied synchronously after every game.
  Metrics are logged to Weights \& Biases and include expected regret statistics, win rates against baseline opponents, information-set update count quantiles, and median policy probabilities.
  The full experiment configuration is shown in Table~\ref{tab:exp-setup}.

  \vspace{0.5em}
  \begin{table}[H]
    \centering
    \small
    \begin{tabular}{l c}
      \toprule
      \textbf{Parameter}                        & \textbf{Value}                     \\
      \midrule
      \multicolumn{2}{l}{\textbf{Game Configuration}}                                \\
      Required property sets                    & 2                                  \\
      Initial hand size                         & 5                                  \\
      New cards per turn                        & 2                                  \\
      Turns per streak                          & 2                                  \\
      \midrule
      \multicolumn{2}{l}{\textbf{Deck Composition}}                                  \\
      Property cards (Brown/Green/Pink)         & 30 (10 each)                       \\
      Cash cards (Value \$1 and \$3)            & 20 (10 each)                       \\
      Rent cards (Brown/Green/Pink)             & 30 (10 each)                       \\
      Special cards (Just Say No)               & 3                                  \\
      Total cards                               & 83                                 \\
      \midrule
      \multicolumn{2}{l}{\textbf{Training Parameters}}                               \\
      Abstraction class                         & \texttt{IntentStateAbstraction}    \\
      Resolver class                            & \texttt{GreedyActionResolver}      \\
      Training games                            & $1,000$                            \\
      Test games per evaluation                 & 20                                 \\
      Evaluation interval                       & 50~games                           \\
      \texttt{RiskAwareSelector} aggressiveness & 0.5                                \\
      \texttt{RiskAwareSelector} temperature    & 2                                  \\
      Simulations per action                    & 20                                 \\
      Buffer size                               & 10                                 \\
      Exploration rate ($\epsilon$)             & 0.1                                \\
      Maximum turns per game                    & 250                                \\
      CPUs used                                 & 8                                  \\
      Parallelism mode                          & \texttt{parallel-unordered-update} \\
      Random seed                               & 1                                  \\
      \bottomrule
    \end{tabular}
    \caption{Training configuration for all experiments.}
    \label{tab:exp-setup}
  \end{table}

  \subsection{Regret Convergence}

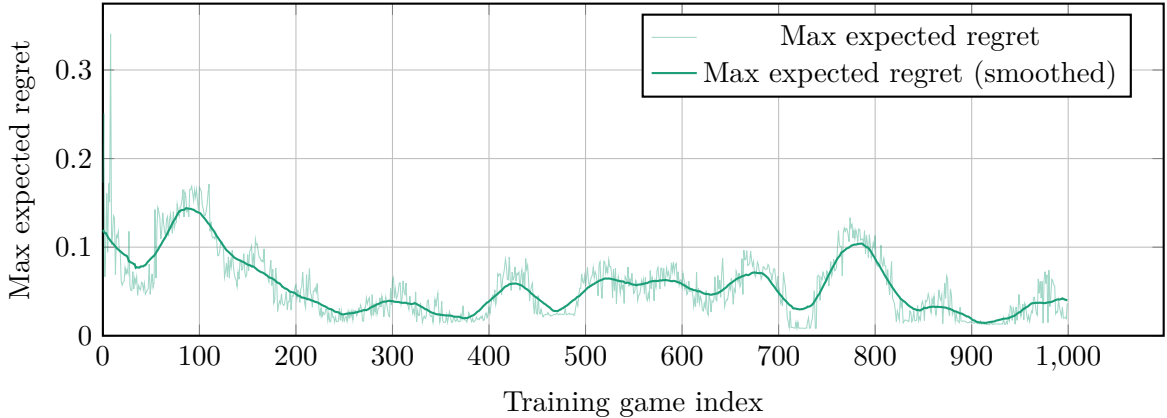
\begin{figure}[H]
    \centering
    \begin{tikzpicture}
      \begin{axis}[
          width=0.95\linewidth,
          height=6cm,
          title={Max Expected Regret During Training},
          xlabel={Training game index},
          ylabel={Max expected regret},
          grid=both,
          xmin=0,
          ymin=0,
          legend pos=north east,
          thick,
        ]
        \addplot+[mark=none, color=cbteal, opacity=0.4, thin]
        table [x=step, y=max_expected_regret, col sep=comma] {data/max_expected_regret.csv};
        \addlegendentry{Max expected regret}

        \addplot+[mark=none, color=cbteal, opacity=1.0, thick]
        table [x=step, y=max_expected_regret_smooth, col sep=comma] {data/max_expected_regret.csv};
        \addlegendentry{Max expected regret (smoothed)}
      \end{axis}
    \end{tikzpicture}
    \vspace{-0.5em}
    \caption{
      Decline in maximum expected regret during training, demonstrating convergence of MCCFR under bounded one-sided response dynamics.
    }
    \label{fig:regret-convergence}
  \end{figure}

  We measure convergence using the maximum expected regret (MER) as defined in Section~\ref{sec:background}. This metric is standard in CFR literature, as it weights each information set by its relevance to the opponent's strategy, ensuring that regret reduction is focused on frequently reached decision points. Figure~\ref{fig:regret-convergence} shows that regret declines during early training and achieves a relatively stable, low-regret state within $1,000$ games (19 minutes of wall-clock training time).

  \subsection{Win Rate Evaluation}

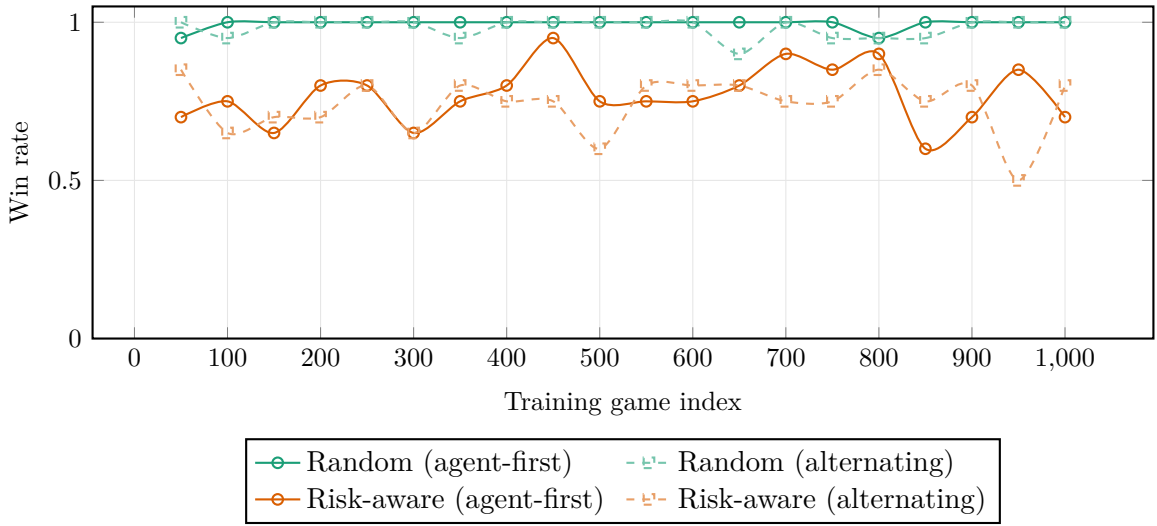
\begin{figure}[H]
    \centering
    \begin{tikzpicture}
      \begin{axis}[
          width=0.95\linewidth,
          height=6cm,
          grid=both,
          grid style={gray!20},
          title={Win Rate Against Baseline Opponents},
          xlabel={Training game index},
          ylabel={Win rate},
          ymin=0, ymax=1.05,
          legend style={
              at={(0.5,-0.3)},
              anchor=north,
              legend columns=2,
              /tikz/every even column/.append style={column sep=0.2cm}
            },
          legend cell align={left},
          thick,
          tick label style={font=\small},
          label style={font=\small},
        ]
        \addplot+[smooth, thick, color=cbteal, solid, mark=o]
        table [x=step, y=random_agent_first, col sep=comma] {data/winrates.csv};
        \addlegendentry{Random (agent-first)}
        \addplot+[smooth, thick, color=cbteal!60!white, dashed, mark=square]
        table [x=step, y=random_alternating, col sep=comma] {data/winrates.csv};
        \addlegendentry{Random (alternating)}
        \addplot+[smooth, thick, color=cborange, solid, mark=o]
        table [x=step, y=risk_agent_first, col sep=comma] {data/winrates.csv};
        \addlegendentry{Risk-aware (agent-first)}
        \addplot+[smooth, thick, color=cborange!60!white, dashed, mark=square]
        table [x=step, y=risk_alternating, col sep=comma] {data/winrates.csv};
        \addlegendentry{Risk-aware (alternating)}
      \end{axis}
    \end{tikzpicture}
    \vspace{-0.5em}
    \caption{
      Win rate over time against baseline opponents (20 games played at each evaluation) during CFR training.
      Solid lines correspond to games where the agent always plays first; dashed lines indicate randomized starts.
      The model achieves near-perfect play against random opponents and competitive win~rates against more sophisticated opponents.
    }
    \label{fig:winrates}
  \end{figure}

  To assess empirical performance, the trained agent was evaluated against two fixed baselines: a \emph{RandomSelector}, which samples uniformly from all legal actions, and a \emph{RiskAwareSelector}, which biases toward property- or cash-oriented actions according to fixed \emph{aggressiveness} and \emph{temperature} parameters. Each baseline was tested under two initialization conditions: \emph{agent-first}, where the agent always plays first, and \emph{alternating-start}, where the starting player is randomized. Figure~\ref{fig:winrates} shows that the agent converges to a nearly 100\%~win~rate against the random baseline and roughly 75\%~win~rate against the risk-aware baseline.

  \subsection{Update Dynamics}

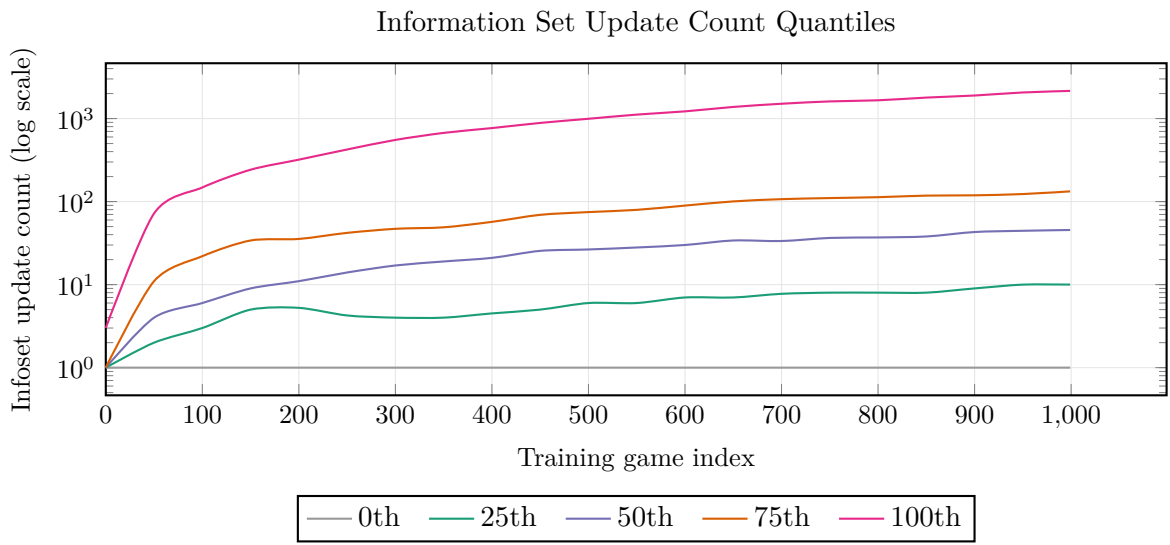
\begin{figure}[H]
    \centering
    \begin{tikzpicture}
      \begin{axis}[
          width=0.95\linewidth,
          height=6cm,
          grid=major,
          grid style={gray!20},
          title={Information Set Update Count Quantiles},
          xlabel={Training game index},
          ylabel={Infoset update count (log scale)},
          ymode=log,
          xmin=0,
          legend style={
              at={(0.5,-0.3)},
              anchor=north,
              legend columns=5,
              /tikz/every even column/.append style={column sep=0.3cm}
            },
          legend cell align={left},
          thick,
          tick label style={font=\small},
          label style={font=\small},
        ]

        \addplot+[smooth, thick, color=cbgray, mark=none]
        table [x=step, y=update_counts_p0, col sep=comma]
          {data/update_counts.csv};
        \addlegendentry{0th}

        \addplot+[smooth, thick, color=cbteal, mark=none]
        table [x=step, y=update_counts_p25, col sep=comma]
          {data/update_counts.csv};
        \addlegendentry{25th}

        \addplot+[smooth, thick, color=cbpurple, mark=none]
        table [x=step, y=update_counts_p50, col sep=comma]
          {data/update_counts.csv};
        \addlegendentry{50th}

        \addplot+[smooth, thick, color=cborange, mark=none]
        table [x=step, y=update_counts_p75, col sep=comma]
          {data/update_counts.csv};
        \addlegendentry{75th}

        \addplot+[smooth, thick, color=cbpink, mark=none]
        table [x=step, y=update_counts_p100, col sep=comma]
          {data/update_counts.csv};
        \addlegendentry{100th}

      \end{axis}
    \end{tikzpicture}

    \vspace{-0.5em}
    \caption{
      Distribution of cumulative infoset update counts throughout training on a logarithmic scale.
      Typical information sets (50th percentile) are visited roughly 50 times during training, while the most common information sets (100th percentile) are visited roughly 2,000 times.
    }
    \label{fig:update-dynamics}
  \end{figure}

  To evaluate how the agent refines its policy, we track the cumulative number of updates per information set throughout training.
  Figure~\ref{fig:update-dynamics} shows the quantiles of these update counts on a logarithmic scale.
  The median update count is roughly 50, while the maximum update count is approximately 2,000.

  \subsection{Policy Evolution}

\begin{figure}[H]
    \centering
    \begin{tikzpicture}
      \begin{axis}[
          width=0.95\linewidth,
          height=6cm,
          grid=both,
          grid style={gray!20},
          title={Median Action Probabilities Throughout Training},
          xlabel={Training game index},
          ylabel={Median action probability},
          ymin=0, ymax=1.05,
          xmin=0,
          legend style={
              at={(0.5,-0.3)},
              anchor=north,
              legend columns=3,
              /tikz/every even column/.append style={column sep=0.2cm}
            },
          legend cell align={left},
          thick,
          tick label style={font=\small},
          label style={font=\small},
        ]

        \addplot+[smooth, thick, color=cbblue] table [x=step, y=add_to_property_set, col sep=comma] {data/median_action_probs.csv};
        \addlegendentry{Add to Property Set}

        \addplot+[smooth, thick, color=cbteal] table [x=step, y=attempt_collect_rent, col sep=comma] {data/median_action_probs.csv};
        \addlegendentry{Attempt Collect Rent}

        \addplot+[smooth, thick, color=cborange] table [x=step, y=cash, col sep=comma] {data/median_action_probs.csv};
        \addlegendentry{Cash}

        \addplot+[smooth, thick, color=cbpurple] table [x=step, y=complete_property_set, col sep=comma] {data/median_action_probs.csv};
        \addlegendentry{Complete Property Set}

        \addplot+[smooth, thick, color=cbpink] table [x=step, y=give_opponent_cash, col sep=comma] {data/median_action_probs.csv};
        \addlegendentry{Give Opponent Cash}

        \addplot+[smooth, thick, color=cbbrown] table [x=step, y=give_opponent_property, col sep=comma] {data/median_action_probs.csv};
        \addlegendentry{Give Opponent Property}

        \addplot+[smooth, thick, color=cbsky] table [x=step, y=just_say_no, col sep=comma] {data/median_action_probs.csv};
        \addlegendentry{Just Say No}

        \addplot+[smooth, thick, color=cbyellow] table [x=step, y=start_new_property_set, col sep=comma] {data/median_action_probs.csv};
        \addlegendentry{Start New Property Set}

        \addplot+[smooth, thick, color=cbyellow] table [x=step, y=pass, col sep=comma] {data/median_action_probs.csv};
        \addlegendentry{Pass}

      \end{axis}
    \end{tikzpicture}

    \vspace{-0.5em}
    \caption{
      Median probability of abstract actions available to the target player throughout training. Actions that promote property building and retention are favored.
    }
    \label{fig:median_action_probs}
  \end{figure}
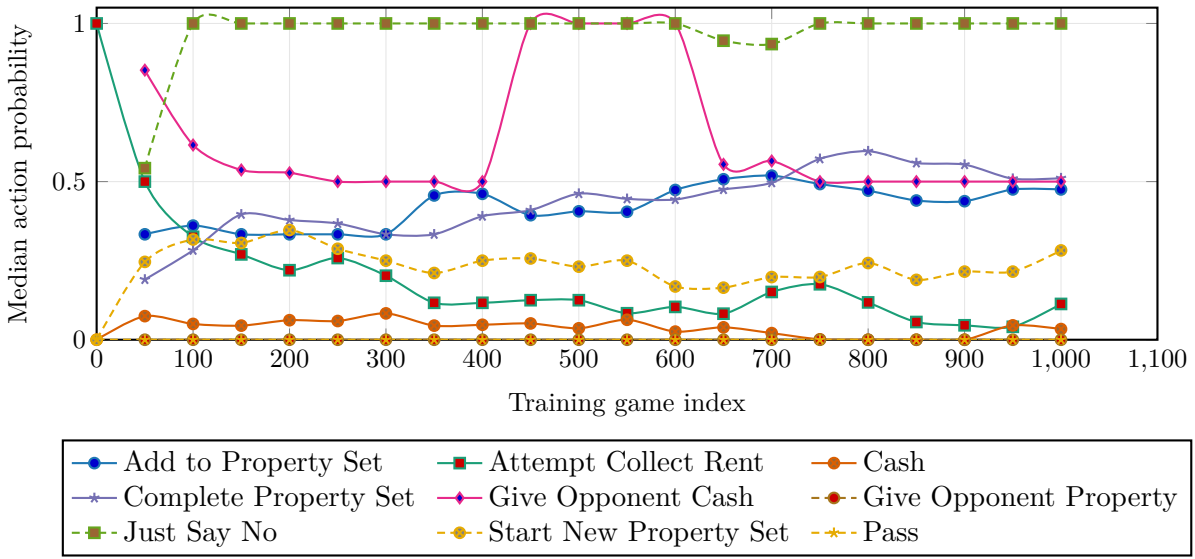

  We next examine the evolution of the learned policy by tracking the median probability of each abstract action across information sets in which it has nonzero probability mass. Intuitively, agents favor \textsc{JustSayNo} and \textsc{GiveOpponentCash} actions over \textsc{GiveOpponentProperty} during the response phase, and \textsc{AddToPropertySet} and \textsc{CompletePropertySet} actions over \textsc{Cash} during normal play. Figure~\ref{fig:median_action_probs} shows the evolution of these probabilities.

  \section{Discussion}

  The experimental results validate the core hypothesis: that a practical, high-performing policy can be derived from a minimal intent-based abstraction using Monte Carlo Counterfactual Regret Minimization in a bounded one-sided response environment. Within $1,000$ games (approximately 19 minutes of wall-clock training), the agent achieves competitive play against strong baselines, validating the soundness of the abstraction and the efficiency of the training runtime.

  Methodologically, the results underscore the value of the $\boldsymbol{\varrho}$-operator for formalizing the BORG dynamic. This modeling enables turn-interrupting, yet finite, interactions to be integrated into the game state without requiring algorithmic modifications to the standard CFR framework. Its structure provides a valuable intermediate testbed for sequential decision-making, bridging the gap between purely sequential environments and fully reciprocal, recursive games.

Finally, we address a key limitation of the current BORG formulation. The agent's turn-interrupting responses are functionally a multi-set decision, meaning the order of actions taken within the response phase does not affect the final outcome. The agent's task is thus equivalent to selecting the optimal combination of cards to play. While this simplifies the initial convergence problem, addressing this multi-set constraint is necessary to fully realize a sequential decision process within the response phase itself.

\section{Conclusion}

We presented a lightweight, full-stack system for studying Bounded One-Sided Response Games (BORGs) through a modified version of \emph{Monopoly Deal}. Our work formalizes the BORG dynamic and demonstrates its tractability, showing that classical Monte Carlo CFR techniques apply directly and converge reliably on modest hardware.

Beyond the empirical results, the platform itself constitutes a key contribution. By integrating the environment, a parallelized CFR runtime, and a web interface into a single, accessible, and highly transparent framework, we offer a system that prioritizes reproducibility and introspection for future research.

\paragraph{Future Work}

Future work will focus on studying sequential decision-making in the response phase itself, using a higher-fidelity state space and modern reinforcement learning techniques.

\begin{enumerate}
  \item \textbf{Sequential Response Dependencies:} An important next step is to introduce action dependencies within the response phase to move beyond the current multi-set decision structure.
  \item \textbf{Policy Generalization and Granularity:} We plan to move beyond tabular methods to modern reinforcement learning techniques for imperfect-information settings, enabling us to learn policies over larger, more granular state spaces and potentially remove the intent-based state abstraction module entirely.
  \item \textbf{Distributed Training:} As policy complexity increases, training deep neural networks using cloud resources may become necessary. This would require integrating distributed training functionality into our framework while upholding the modularity and introspection guarantees of our current design.
\end{enumerate}
Together, these extensions will broaden the platform's utility as a foundation for practical research in bounded one-sided response games.

\printbibliography

\appendix
\section*{Appendix}

\section{Information Set Key Structure and Serialization}

The system relies on a consistent, deterministic identifier for every unique decision point encountered during training, known as the \emph{information set} (InfoSet) key. This key facilitates state recovery and policy lookup.

The InfoSet key is structured as a unique composite string following the format:
$$
  \texttt{\{player\_idx\}@\{abstraction\_cls\_name\}@\{abstraction\_key\}}
$$
\noindent
\textbf{Serialization Process.} The key is generated through a three-step process: 1) The game state is mapped by the chosen \texttt{abstraction\_cls} to a compact, abstract state (e.g., a tuple of available abstract actions and the streak index), 2) This abstract state is deterministically serialized into a canonical JSON string, and 3) An \texttt{MD5} hash of the resulting JSON string is computed to create the final \texttt{abstraction\_key}.

This process guarantees that any time a player reaches a specific decision point, the resulting \texttt{InfoSet} key is identical.

\section{Baseline Opponent Policies}

To provide empirical validation for the trained CFR agent's performance, the model is evaluated against two fixed, non-learning baseline policies. These baselines establish a lower bound and a reasonable heuristic benchmark for strategic play (Section~\ref{sec:experiments}). Both models operate using the same abstract action space and choose a concrete action via the deterministic \texttt{GreedyActionResolver}.

\begin{enumerate}
  \item \textbf{Random Selector (\texttt{RandomSelector}):} This policy serves as the statistical baseline, sampling uniformly from the entire set of legal abstract actions available at any given information set.
  \item \textbf{Risk-Aware Selector (\texttt{RiskAwareSelector}):} This policy implements a simple heuristic strategy by balancing a fixed \emph{aggressiveness} parameter (ranging from 0 to 1). The policy biases its action choice toward prioritizing property acquisition (aggressiveness) over banking cash (caution), using a \emph{temperature} parameter to control the sensitivity of this preference. This model exhibits a fixed, consistent strategy that is more competitive than pure random play.
\end{enumerate}

During evaluation, the CFR agent is measured against both baselines under two starting conditions: always going first (\emph{agent-first}) and alternating who starts the game (\emph{alternating-start}).

\section{Checkpoint Structure}
\label{sec:checkpoint}

To ensure full reproducibility and auditability, the system serializes the complete learner state after a configurable number of games into a single, human-readable JSON checkpoint. The checkpoint acts as a single source of truth, enabling an experiment to be resumed or played back exactly from any saved index.

The checkpoint is composed of three primary nested components that store the complete CFR state, alongside a metadata section:

\begin{enumerate}
  \item \textbf{Game Configuration and Metadata:} Contains all static parameters defining the environment and training run, including the deck configuration, abstraction class, resolver class, simulations per action, exploration $\epsilon$, and more.
  \item \textbf{Policy Manager:} Maintains a rolling buffer of recent policy vectors for each player, tracking the buffer capacity and the number of times each information set has been updated. Once deserialized, this data structure maps information set keys to average and ``current'' probability vectors over abstract actions.
  \item \textbf{Regret Manager:} Stores the total regret and visit counts for all abstract actions (represented by their numerical indices) in every encountered information set. Once deserialized, this data structure maps information set keys and available abstract actions to sum and mean regret values.
  \item \textbf{Counterfactual Reach Probability Counter:} Tracks the counterfactual reach probability of each information set encountered during self-play. Presently, these values are used to compute the maximum expected regret during training.
\end{enumerate}

An abridged example of the checkpoint structure is shown in Figure \ref{lst:checkpoint}.

\begin{figure}[H]
  \centering
  \begin{verbatim}
{
    "game_config": {
        "cash_card_values": [1,3],
        "required_property_sets": 2,
        "deck_size_multiplier": 5,
        "initial_hand_size": 5,
        "...": "..."
    },
    "abstraction_cls": "IntentStateAbstraction",
    "resolver_cls": "GreedyActionResolver",
    "sims_per_action": 50,
    "epsilon": 0.1,
    "policy_manager": {
        "buffer_size": 10,
        "update_count": {
            "1@IntentStateAbstraction@d1daef9db...": 1,
            "...": "..."
        },
        "player_buffers": [
            {
                "buffer": {
                    "0@IntentStateAbstraction@c434147c...": [
                        [1.0, 0.0, 0.0, ...]
                    ],
                    "...": "..."
                },
                "buffer_size": 10
            }
        ]
    },
    "regret_manager": {
        "1@IntentStateAbstraction@d1daef9db...": {
            "0": {"sum": 0.22, "n": 1.0},
            "8": {"sum": -0.22, "n": 1.0},
            "...": "..."
        },
        "...": "..."
    },
    "cf_reach_prob_counter": {
        "1@IntentStateAbstraction@d1daef9db...": 0.00012,
        "...": "..."
    }
}
\end{verbatim}
  \caption{Abridged structure of the JSON checkpoint, showing the three primary components (\texttt{policy\_manager}, \texttt{regret\_manager}, \texttt{cf\_reach\_prob\_counter}) and core configuration metadata.}
  \label{lst:checkpoint}
\end{figure}

\section{Interactive Web Interface}
\label{sec:ui_screenshot}

\begin{figure}[h]
  \centering
  \includegraphics[width=0.95\linewidth]{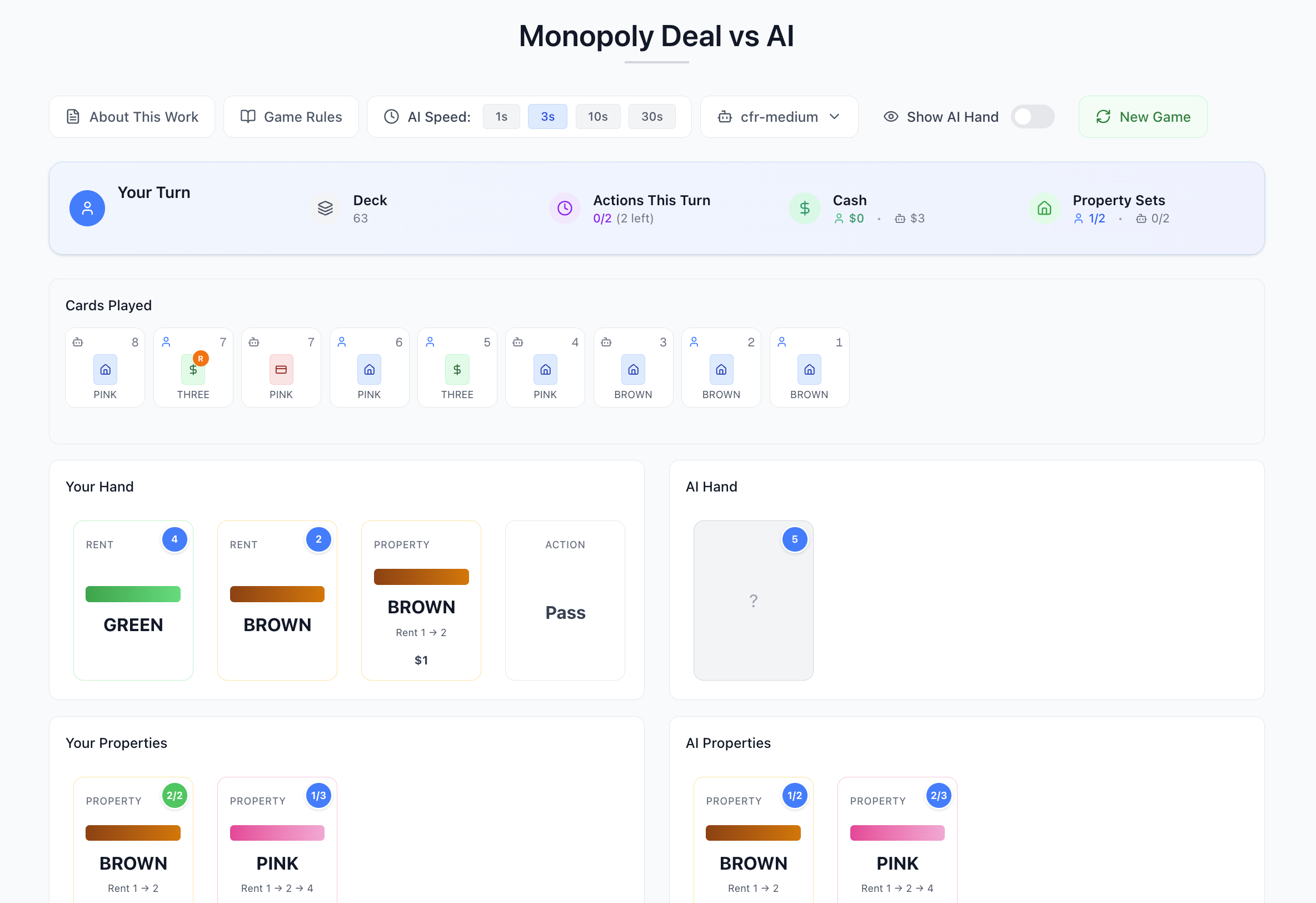}
  \caption{
    Screenshot of the interactive web interface. The interface renders the game state in real time, displays the history of actions, and allows a human player to compete against the trained CFR agent.
  }
  \label{fig:ui_screenshot}
\end{figure}

A web application is provided for researchers to play against trained CFR policies. Built on a FastAPI backend and a Next.js frontend, this interface bridges quantitative training results with qualitative behavioral analysis (Figure \ref{fig:ui_screenshot}).

\end{document}